\documentclass[10pt]{iopartmod}
\usepackage{amssymb}
\usepackage{amsmath}
\usepackage{makeidx}
\usepackage[dvips]{graphicx}

\input{tcilatex}

\begin{document}

\title{Frozen light in periodic metamaterials}

\author{Alex Figotin and Ilya Vitebskiy}

\begin{abstract}
Wave propagation in spatially periodic media, such as photonic crystals, can
be qualitatively different from any uniform substance. The differences are
particularly pronounced when the electromagnetic wavelength is comparable to
the primitive translation of the periodic structure. In such a case, the
periodic medium cannot be assigned any meaningful refractive index. Still,
such features as negative refraction and/or opposite phase and group
velocities for certain directions of light propagation can be found in almost
any photonic crystal. The only reservation is that unlike hypothetical uniform
left-handed media, photonic crystals are essentially anisotropic at frequency
range of interest. Consider now a plane wave incident on a semi-infinite
photonic crystal. One can assume, for instance, that in the case of positive
refraction, the normal components of the group and the phase velocities of the
transmitted Bloch wave have the same sign, while in the case of negative
refraction, those components have opposite signs. What happens if the normal
component of the transmitted wave group velocity vanishes? Let us call it a
"zero-refraction" case. At first sight, zero normal component of the
transmitted wave group velocity implies total reflection of the incident wave.
But we demonstrate that total reflection is not the only possibility. Instead,
the transmitted wave can appear in the form of an abnormal grazing mode with
huge amplitude and nearly tangential group velocity. This spectacular
phenomenon is extremely sensitive to the frequency and direction of
propagation of the incident plane wave. These features can be very attractive
in numerous applications, such as higher harmonic generation and wave mixing,
light amplification and lasing, highly efficient superprizms, etc.

\end{abstract}
\affiliation{University of California at Irvine}
\maketitle

\section{Introduction}

Wave propagation in spatially periodic media, such as photonic crystals, is
qualitatively different from that of any uniform substance. The differences
are particularly pronounced when the wavelength is comparable to the primitive
translation $L$ of the periodic structure
\cite{Joann,Strat1,Strat3,Brill,Notomi,LLEM,Yariv}. The effects of strong
spatial dispersion culminate when the group velocity of a traveling Bloch wave
vanishes. One reason for this is that vanishing group velocity always implies
a dramatic increase in density of modes at the respective frequency. In
addition, vanishing group velocity also implies certain qualitative changes in
the eigenmode structure, which can be accompanied by some spectacular effects
in electromagnetic wave propagation. A particular example of the kind is the
frozen mode regime associated with a dramatic enhancement of the wave
transmitted to the periodic medium \cite{PRE01,PRB03,PRE03,PRE05,JMMM06,WRM06}%
. In this paper we consider several different modifications of the frozen mode
regime, each related to a specific singularity of the electromagnetic
dispersion relation. For every case we present specific examples of periodic
dielectric structures supporting such a phenomenon. For the most part, we
restrict ourselves to periodic layered media, which are photonic crystals with
one dimensional periodicity.

\begin{figure}[tbph]
\scalebox{0.8}{\includegraphics[viewport=0 0 500 180,clip]{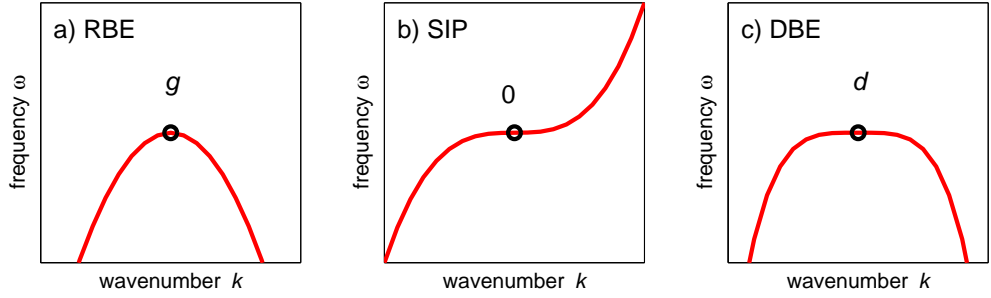}}
\caption{Fragments of dispersion relations containing different stationary
points: (a) a regular band edge (RBE) $g$, (b) a stationary inflection appoint
(SIP) $0$, (c) a degenerate band edge (DBE) $d$.}
\label{DRSP3}
\end{figure}

\begin{figure}[tbph]
\scalebox{0.8}{\includegraphics[viewport=-50 0 500 200,clip]{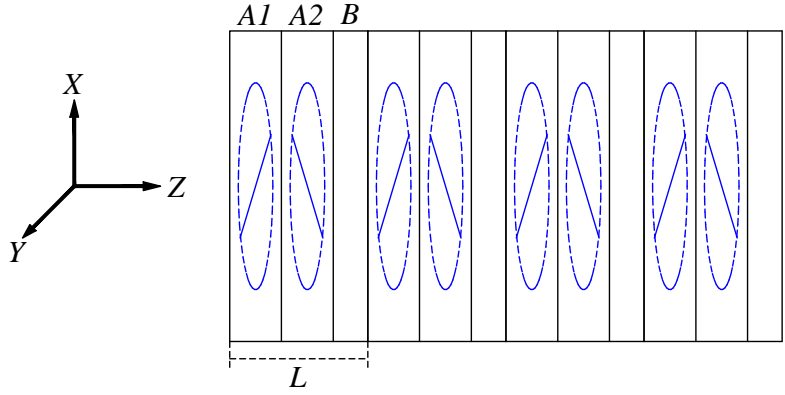}}
\caption{Periodic stack capable of supporting $k-\omega$ diagram with
degenerate band edge. A unit cell $L$ includes three layers: two birefringent
layers $A_{1}$ and $A_{2}$ with misaligned in-plane anisotropy and one
isotropic $B$ layer. The misalignment angle $\varphi=\varphi_{1}-\varphi_{2}$
between adjacent anisotropic layers $A_{1}$ and $A_{2}$ must be different from
$0$ and $\pi/2$.}
\label{StackAAB}
\end{figure}

We start with the simple case of a plane electromagnetic wave normally
incident on a semi-infinite periodic layered structure. In the case of normal
incidence, there are two different kinds of frozen mode regime. The first one
is associated with a stationary inflection point of the dispersion relation in
Fig. \ref{DRSP3}(b). In this case the incident plane wave of frequency
$\omega_{0}$ is converted into the frozen mode with infinitesimal group
velocity and very large diverging amplitude. A detailed description of this
phenomenon can be found in our previous publications on this subject
\cite{PRE01,PRB03,PRE03,PRE05,JMMM06,WRM06}. Qualitatively different kind of
frozen mode regime occurs in the vicinity of a degenerate photonic band edge,
shown in Fig. \ref{DRSP3}(c). In this latter case, the incident wave of
frequency $\omega_{d}$ is reflected back to space, but not before it creates a
frozen mode with huge diverging amplitude. Not every periodic structure can
support the frozen mode regime. Specifically, in the case of a normal
incident, a unit cell of the periodic layered array must contain at least
three layers, of which two must display a misaligned in-plane anisotropy, as
shown in Fig. \ref{StackAAB}. In photonic crystals with three dimensional
periodicity, the presence of anisotropic constitutive component is not necessary.

\begin{figure}[tbph]
\scalebox{0.8}{\includegraphics[viewport=-50 0 500 220,clip]{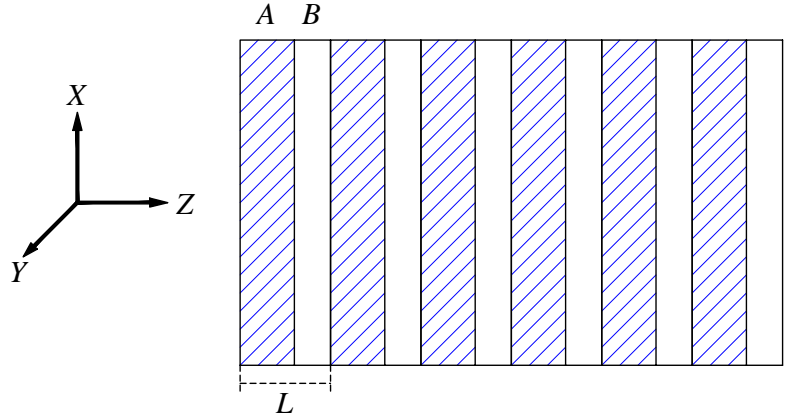}}
\caption{Periodic layered structure with two layers $A$ and $B$ in a primitive
cell $L$. The $A$ layers (hatched) are anisotropic with one of the principle
axes of the dielectric permittivity tensor making an oblique angle with the
normal $z$ to the layers ($\varepsilon_{xz}\neq0$). The $B$ layers are
isotropic. The $x-z$ plane coincides with the mirror plane of the stack.}
\label{StackAB}
\end{figure}

\begin{figure}[tbph]
\scalebox{0.8}{\includegraphics[viewport=-50 0 500 200,clip]{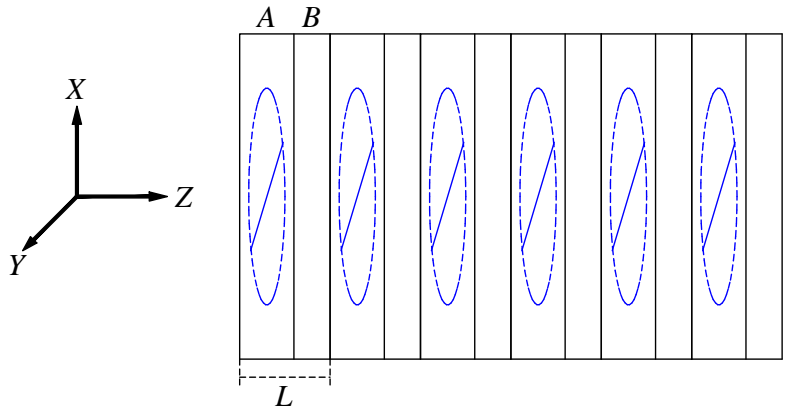}}
\caption{Periodic layered structure with two layers $A$ and $B$ in a unit cell
$L$. The $A$ layer has inplane anisotropy, while the $B$ layer can be
isotropic.}
\label{StackABc}
\end{figure}

In Section 3, we turn to the case of oblique wave propagation, which is of
particular interest here. The frozen mode regime at oblique incidence can
occur when the normal component of the transmitted wave group velocity
vanishes, while its tangential component remains finite. In such a case, the
transmitted wave is an abnormal grazing mode with a dramatically enhanced
amplitude and nearly tangential energy flux. Remarkably, the photonic crystal
reflectivity at this point can be very low, implying almost total conversion
of the incident plane wave into the abnormal grazing mode. A significant
advantage of this phenomenon is that it can occur in much simpler periodic
structures, compared to those supporting the frozen mode regime at normal
incidence. Examples are shown in Fig. \ref{StackAB} and \ref{StackABc}. The
presence of anisotropic layers is still required.

Yet another interesting modification of the frozen mode regime is an abnormal
subsurface wave in the vicinity of a degenerate photonic band edge. A regular
surface wave usually decays exponentially with the distance from the surface
in either direction. By contrast, the abnormal subsurface wave is extremely
asymmetric. It does decay rapidly outside the photonic crystal. But inside the
periodic medium, its amplitude sharply increases, and reaches its maximum at a
certain distance from the surface. Only after that the field amplitude begins
a slow decay, as the distance from the surface further increases. The profile
of such a subsurface wave is similar to that of the frozen mode in the
vicinity of a degenerate band edge. This phenomenon is briefly discussed in
section 4.%

\section{Frozen mode regime at normal incidence}

\begin{figure}[tbph]
\scalebox{0.8}{\includegraphics[viewport=-100 0 500 250,clip]{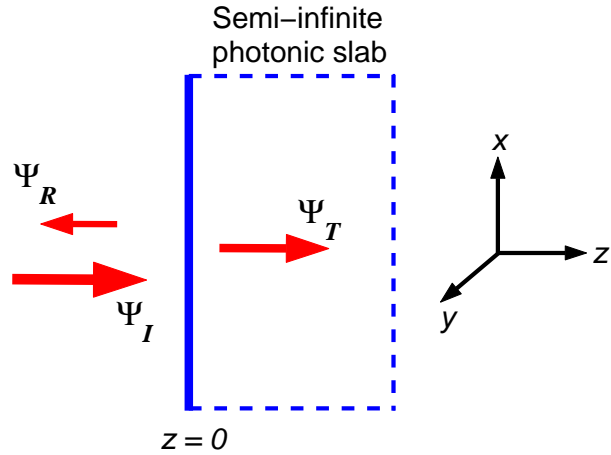}}
\caption{Plane wave normally incident on a semi-infinite photonic crystal. The
subscripts $I$, $R$, and $T$ refer to the incident, reflected and transmitted
waves, respectively. In all cases, the amplitude of the incident wave is
unity.}
\label{SISn}
\end{figure}

The essence of the frozen mode regime can be understood from the example of a
plane monochromatic wave normally incident on a semi-infinite periodic layered
structure, as shown in Fig. \ref{SISn}. An important requirement, though, is
that some of the layers must display a misaligned in-plane anisotropy. The
simplest periodic layered array of the kind is shown in Fig. \ref{StackAAB}.

To start with, let us introduce some basic notations and definitions. Let
$\Psi_{I}$, $\Psi_{R}$, and $\Psi_{T}$ be the incident, reflected and
transmitted waves, respectively. Assume for now that all three monochromatic
waves propagate along the $z$ axis normal to the boundary of semi-infinite
periodic layered structure in Figs. \ref{SISn}. The photonic crystal boundary
is located at $z=0$. Electromagnetic field both inside (at $z>0$) and outside
(at $z<0$) the periodic stack is independent of the $x$ and $y$ coordinates,
and its transverse components can be represented as a column-vector
\begin{equation}
\Psi\left(  z\right)  =\left[
\begin{array}
[c]{c}%
E_{x}\left(  z\right) \\
E_{y}\left(  z\right) \\
H_{x}\left(  z\right) \\
H_{y}\left(  z\right)
\end{array}
\right]  , \label{Psi}%
\end{equation}
where $\vec{E}\left(  z\right)  $ and $\vec{H}\left(  z\right)  $ are
time-harmonic electric and magnetic fields. We also assume that anisotropic
layers of the periodic array have in-plane anisotropy, so the fields $\vec
{E}\left(  z\right)  $ and $\vec{H}\left(  z\right)  $ are normal to the
direction of propagation%
\begin{equation}
\vec{E}\left(  z\right)  \perp z,\vec{H}\left(  z\right)  \perp z.
\label{Ez=Hz=0}%
\end{equation}
All four transverse field components in (\ref{Psi}) are continuous functions
of $z$, which leads to the following standard boundary condition at $z=0$
\begin{equation}
\Psi_{T}\left(  0\right)  =\Psi_{I}\left(  0\right)  +\Psi_{R}\left(
0\right)  . \label{BC}%
\end{equation}
Note that the polarizations of the incident, reflected and transmitted waves
can be different, because some of the layers of the periodic array display an
in-plane anisotropy, as shown in the example in Fig. \ref{StackAAB}. The
latter is essential for the possibility of the frozen mode regime.

In periodic layered media, the electromagnetic eigenmodes $\Psi_{k}\left(
z\right)  $ are usually chosen in the Bloch form%
\begin{equation}
\Psi_{k}\left(  z+L\right)  =e^{ikL}\Psi_{k}\left(  z\right)  , \label{BF}%
\end{equation}
where the Bloch wavenumber $k$ is defined up to a multiple of $2\pi/L$. The
correspondence between $\omega$ and $k$ is referred to as the Bloch dispersion
relation. Real $k$ correspond to propagating (traveling) Bloch modes.
Propagating modes belong to different spectral branches $\omega\left(
k\right)  $ separated by frequency gaps. The speed of a traveling wave in
periodic medium is determined by the group velocity \cite{Brill}%
\begin{equation}
u=d\omega/dk. \label{u}%
\end{equation}
Normally, each spectral branch $\omega\left(  k\right)  $ develops stationary
points $\omega_{s}=\omega\left(  k_{s}\right)  $ where the group velocity
(\ref{u}) of the corresponding propagating mode vanishes%
\begin{equation}
d\omega/dk=0\text{, at }\omega=\omega_{s}=\omega\left(  k_{s}\right)  .\text{
} \label{SP}%
\end{equation}
Examples of different stationary points are shown in Fig. \ref{DRSP3}, where
each of the frequencies $\omega_{g}$, $\omega_{0}$ and $\omega_{d}$ is
associated with zero group velocity of the respective traveling wave.
Stationary points (\ref{u}) play essential role in the formation of frozen
mode regime.

By contrast, evanescent Bloch modes are characterized by complex wavenumbers
$k=k^{\prime}+ik^{\prime\prime}$. Evanescent modes decay exponentially with
the distance $z$ from the boundary of semi-infinite periodic structure.
Therefore, under normal circumstances, evanescent contribution to the
transmitted wave $\Psi_{T}\left(  z\right)  $ can be significant only in close
proximity of the surface. The situation can change dramatically when the
frequency $\omega$ approaches one of the stationary point values $\omega_{s}$.
At first sight, stationary points (\ref{SP}) relate only to propagating Bloch
modes. But in fact, in the vicinity of every stationary point frequency
$\omega_{s}$, the imaginary part $k^{\prime\prime}$ of the Bloch wavenumber of
at least one of the evanescent modes also vanishes. As a consequence, the
respective evanescent mode decays very slowly, and its role may extend far
beyond the photonic crystal boundary. In addition, in the special cases of
interest, the electromagnetic field distribution$\Psi\left(  z\right)  $ in
the coexisting evanescent and propagating eigenmodes becomes very similar, as
$\omega$ approaches $\omega_{s}$. This can result in spectacular resonance
effects, such as the frozen mode regime. What exactly happens in the vicinity
of a particular stationary point (\ref{SP}) essentially depends on its
character and appears to be very different in each of the three cases
presented in Fig. \ref{DRSP3}. In the next subsection we present a simple
qualitative picture of the frozen mode regime based on energy conservation
consideration. Then, in the following subsection, we provide more consistent
analysis of the phenomenon.%

\subsection{Energy density and energy flux at frozen mode regime}

Let $S_{I}$, $S_{R}$ and $S_{T}$ be the energy flux in the incident, reflected
and transmitted waves in Fig. \ref{SISn}. The transmission and reflection
coefficients of a lossless semi-infinite medium are defined as%
\begin{equation}
\tau=\frac{S_{T}}{S_{I}},\ \rho=-\frac{S_{R}}{S_{I}}, \label{n tau-rho}%
\end{equation}
where%
\[
S_{I}+S_{R}=S_{T},~\rho=1-\tau.
\]
With certain reservations, the energy flux $S_{T}$ of the transmitted
travelling wave can be expressed as%
\begin{equation}
S_{T}=W_{T}u, \label{S=Wu}%
\end{equation}
where $u$ is the group velocity, which is also the energy velocity, and
$W_{T}$ is the energy density%
\[
W_{T}\propto\left\vert \Psi_{T}\right\vert ^{2}.
\]
Evanescent modes do not contribute to the normal energy flux $S_{T}$ in the
case of a lossless semi-infinite periodic structure. Besides, evanescent
contribution to the transmitted wave becomes negligible at a certain distance
$z$ from the surface. The simple expression (\ref{S=Wu}) may not apply when
the transmitted wave involves two or more propagating Bloch modes, but we will
not deal with such a situation here.

Vanishing group velocity $u$ in (\ref{S=Wu}) implies that the transmitted wave
energy flux $S_{T}$ also vanishes, along with the respective transmission
coefficient $\tau$ in (\ref{n tau-rho}). The only exception could be if the
energy density $W_{T}$ of the transmitted wave increases dramatically in the
vicinity of stationary point frequency. In other words, if $W_{T}$ in
(\ref{S=Wu}) grows fast enough, as $\omega$ approaches $\omega_{s}$, the
product $W_{T}u$ in (\ref{S=Wu})\ can remain finite even at $\omega=\omega
_{s}$. In such a case, a significant fraction of the incident radiation can be
converted into the slow mode inside the semi-infinite periodic array.
Hereinafter, the effect of a dramatic growth of the transmitted wave amplitude
in the vicinity of a stationary point (\ref{SP}) will be referred to as the
frozen mode regime. The possibility of such an effect is directly related to
the character of a particular stationary point. From this point of view, let
us consider three different situation presented in Fig. \ref{DRSP3}.

\subsubsection{Regular band edge}

We start with the simplest case of a regular photonic band edge (RBE) shown in
Fig. \ref{DRSP3}(a). It can be found in any periodic array, including any
periodic layered structure. Just below the band edge frequency $\omega_{g}$,
the dispersion relation can be approximated by a quadratic parabola%
\begin{equation}
\omega_{g}-\omega\propto\left(  k-k_{g}\right)  ^{2},\text{ at }%
\omega\lessapprox\omega_{g}. \label{RBE DR}%
\end{equation}
This yields the following frequency dependence of the propagating mode group
velocity $u$ below the photonic band edge
\begin{equation}
u=\frac{d\omega}{dk}\propto\left(  k_{g}-k\right)  \propto\left(  \omega
_{g}-\omega\right)  ^{1/2},\text{ at }\omega\lessapprox\omega_{g}.
\label{u(RBE)}%
\end{equation}
Due to the boundary condition (\ref{BC}), the amplitude of the transmitted
propagating Bloch mode in this case remains finite and comparable to that of
the incident wave. Therefore, the energy flux (\ref{S=Wu}) associated with the
transmitted slow mode vanishes, as $\omega$ approaches $\omega_{g}$%
\begin{equation}
S_{T}=W_{T}u\propto\left\{
\begin{array}
[c]{c}%
\left(  \omega_{g}-\omega\right)  ^{1/2},\text{ at }\omega\lessapprox
\omega_{g}\\
0,\text{ at }\omega\geq\omega_{g}%
\end{array}
\right.  . \label{ST(RBE)}%
\end{equation}
Formula (\ref{ST(RBE)}) expresses the well-known fact that in the vicinity of
a regular band edge, a lossless semi-infinite photonic crystal becomes totally reflective.

\subsubsection{Stationary inflection point}

A completely different situation occurs in the vicinity of a stationary
inflection point (SIP) shown in Fig. \ref{DRSP3}(b). Such a point can be found
in periodic layered structures involving anisotropic and magnetic layers
\cite{PRE01,PRB03}, as well as in some photonic crystals with 2- and
3-dimensional periodicity. In the vicinity of a stationary inflection point
$\omega_{0}$, the dispersion relation can be approximated by a cubic parabola%
\begin{equation}
\omega-\omega_{0}\propto\left(  k-k_{0}\right)  ^{3}. \label{SIP DR}%
\end{equation}
The propagating mode group velocity $u$ vanishes as $\omega$ approaches
$\omega_{0}$ from either direction%
\begin{equation}
u=\frac{d\omega}{dk}\propto\left(  k-k_{0}\right)  ^{2}\propto\left(
\omega-\omega_{0}\right)  ^{2/3}. \label{u(SIP)}%
\end{equation}
But remarkably, the amplitude of the transmitted propagating mode increases so
that the respective energy density $W_{T}$ diverges as $\omega\rightarrow
\omega_{0}$%
\begin{equation}
W_{T}\propto\left(  \omega-\omega_{0}\right)  ^{-2/3}. \label{W(SIP)}%
\end{equation}
The expression (\ref{u(SIP)}) together with (\ref{W(SIP)}) yield that the
energy flux of the transmitted slow mode remains finite even at $\omega
=\omega_{0}$%
\begin{equation}
S_{T}=W_{T}u\sim S_{I},\text{ at }\omega\approx\omega_{0}. \label{ST(SIP)}%
\end{equation}
The latter implies that the incident light is converted to the frozen mode
with infinitesimal group velocity (\ref{u(SIP)}) and huge diverging amplitude
(\ref{W(SIP)}). Of course, in reality, the amplitude of the frozen mode will
be limited by such factors as absorption, nonlinear effects, imperfection of
the periodic dielectric array, deviation of the incident radiation from a
perfect plane monochromatic wave, finiteness of the photonic crystal
dimensions, etc. Still, with all these limitations in place, the frozen mode
regime can be very attractive for a variety of practical applications.

\subsubsection{Degenerate band edge}

Let us turn to the case of a degenerate band edge (DBE) shown in Fig.
\ref{DRSP3}(c). Such a point can be found in dispersion relation of periodic
layered structures involving misaligned anisotropic layers. An example is
shown in Fig. \ref{StackAAB}. Just below the degenerate band edge frequency
$\omega_{d}$, the dispersion relation $\omega\left(  k\right)  $ can be
approximated as%
\begin{equation}
\omega_{d}-\omega\propto\left(  k-k_{d}\right)  ^{4},\text{ at }%
\omega\lessapprox\omega_{d} \label{DBE DR}%
\end{equation}
This yields the following frequency dependence of the propagating mode group
velocity%
\begin{equation}
u=\frac{d\omega}{dk}\propto\left(  k_{d}-k\right)  ^{3}\propto\left(
\omega_{d}-\omega\right)  ^{3/4},\text{ at }\omega\lessapprox\omega_{d}.
\label{u(DBE)}%
\end{equation}
The amplitude of the transmitted slow mode in this case diverges, as the
frequency approaches the band edge value%
\begin{equation}
W_{T}\propto\left\vert \omega_{d}-\omega\right\vert ^{-1/2},\text{
}\ \ \text{at }\omega\lessapprox\omega_{DBE}, \label{W(DBE)}%
\end{equation}
which constitutes the frozen mode regime. But the amplitude does not grow fast
enough to offset the vanishing group velocity (\ref{u(DBE)}). Indeed, the
expressions (\ref{u(DBE)}) and (\ref{W(DBE)}) together with (\ref{S=Wu}) yield
for the energy flux%
\begin{equation}
S_{T}=W_{T}u\propto\left\{
\begin{array}
[c]{c}%
\left(  \omega_{d}-\omega\right)  ^{1/4},\text{ at }\omega\lessapprox
\omega_{d}\\
0,\text{ at }\omega\geq\omega_{d}%
\end{array}
\right.  , \label{ST(DBE)}%
\end{equation}
implying that, in spite of the diverging energy density (\ref{W(DBE)}), the
energy flux of the transmitted slow wave vanishes, as $\omega$ approaches
$\omega_{d}$.

The situation at a degenerate band edge (\ref{DBE DR}) can be viewed as
intermediate between the frozen mode regime at a stationary inflection point
(\ref{SIP DR}), and the vicinity of a regular band edge (\ref{RBE DR}).
Indeed, on the one hand, the incident wave at $\omega=\omega_{d}$ is totally
reflected back to space, as is the case at a regular band edge. On the other
hand, the transmitted field amplitude inside the periodic medium becomes very
large as $\omega\rightarrow\omega_{d}$, which is similar to what occurs at a
stationary inflection point. The very large amplitude (\ref{W(DBE)}) of the
transmitted wave at $\omega\approx\omega_{d}$ still can be very attractive for applications.

The above consideration does not explain the nature of the frozen mode regime,
nor does it address the problem of the eigenmode composition of the frozen
mode. All these questions are the subject of the next subsection.

\subsection{Physical nature of the frozen mode regime}

\subsubsection{Eigenmode composition of transmitted wave}

In a periodic layered structure, at any given frequency $\omega$, there are
four electromagnetic eigenmodes with different polarizations and wavenumbers.
But in the setting of Fig. \ref{SISn} where the semi-infinite periodic layered
array occupies the half-space $z\geq0$, the transmitted wave is a
superposition of only two of the four Bloch eigenmodes. Indeed, the
propagating waves with negative group velocity do not contribute to $\Psi
_{T}\left(  z\right)  $, nor do the evanescent modes exponentially growing
with the distance $z$. Generally, one can distinguish three different possibilities.

\begin{enumerate}
\item Both Bloch components of the transmitted wave $\Psi_{T}$ are propagating
modes%
\begin{equation}
\Psi_{T}\left(  z\right)  =\Psi_{pr1}\left(  z\right)  +\Psi_{pr2}\left(
z\right)  ,\ \;z\geq0. \label{PsiT=pr+pr}%
\end{equation}
The transmitted wave $\Psi_{T}$ in (\ref{PsiT=pr+pr}) is composed of two Bloch
eigenmodes with two different real wavenumbers $k_{1}$ and $k_{2}$ and two
different group velocities $u_{1}>0$ and $u_{2}>0$. This constitutes the
phenomenon of double refraction, provided that $u_{1}$ and $u_{2}$ are
different. The other two Bloch components of the same frequency have negative
group velocities and cannot contribute to the transmitted wave $\Psi_{T}$.

\item Both Bloch components of $\Psi_{T}$ are evanescent%
\begin{equation}
\Psi_{T}\left(  z\right)  =\Psi_{ev1}\left(  z\right)  +\Psi_{ev2}\left(
z\right)  ,\ \;z\geq0. \label{PsiT=ev+ev}%
\end{equation}
The respective two values of $k$ are complex with positive imaginary parts
$k^{\prime\prime}>0$. This is the case when the frequency $\omega$ falls into
photonic band gap at $\omega>\omega_{g}$ in Fig. \ref{DRSP3}(a) or at
$\omega>\omega_{d}$ in Fig. \ref{DRSP3}(c). The fact that $k^{\prime\prime}>0$
implies that the wave amplitude decays with the distance $z$ from the surface
of the semi-infinite photonic crystal. In the case (\ref{PsiT=ev+ev}), the
incident wave is totally reflected back to space by the semi-infinite periodic structure.

\item One of the Bloch components of the transmitted wave $\Psi_{T}$ is a
propagating mode with $u>0$, while the other is an evanescent mode with
$k^{\prime\prime}>0$%
\begin{equation}
\Psi_{T}\left(  z\right)  =\Psi_{pr}\left(  z\right)  +\Psi_{ev}\left(
z\right)  ,\ \;z\geq0. \label{PsiT=pr+ev}%
\end{equation}
For example, this is the case at $\omega\sim\omega_{0}$ in Fig. \ref{DRSP3}%
(b), as well as at $\omega<\omega_{g}$ in Fig. \ref{DRSP3}(a) and at
$\omega<\omega_{d}$ in Fig. \ref{DRSP3}(c). As the distance $z$ from the
surface increases, the evanescent contribution $\Psi_{ev}$ in
(\ref{PsiT=pr+ev}) decays as $\exp\left(  -zk^{\prime\prime}\right)  $, and
the resulting transmitted wave $\Psi_{T}\left(  z\right)  $ turns into a
single propagating Bloch mode $\Psi_{pr}$.
\end{enumerate}

Propagating modes with $u<0$, as well as evanescent modes with $k^{\prime
\prime}<0$, never contribute to the transmitted wave $\Psi_{T}$ inside the
semi-infinite stack in Fig. \ref{SISn}. This statement is based on the
following two assumptions:

\begin{itemize}
\item[-] The transmitted wave $\Psi_{T}$ and the reflected wave $\Psi_{R}$ are
originated from the plane wave $\Psi_{I}$ incident on the semi-infinite
photonic slab from the left, as shown in Fig. \ref{SISn}.

\item[-] The layered array in Fig. \ref{SISn} occupies the entire half-space
and is perfectly periodic at $z>0$.
\end{itemize}

If either of the above conditions is violated, the electromagnetic field
inside the periodic stack can be a superposition of four Bloch eigenmodes with
either sign of the group velocity $u$ of propagating contributions, or either
sign of $k^{\prime\prime}$\ of evanescent contributions. For example, this
would be the case if the periodic layered array in Fig. \ref{SISn} had some
kind of structural defects or a finite thickness.

The propagating modes with $u>0$ and evanescent modes with $k^{\prime\prime
}>0$ are referred to as \emph{forward} waves. Only forward modes contribute to
the transmitted wave $\Psi_{T}\left(  z\right)  $ in the case of a periodic
semi-infinite stack. The propagating modes with $u<0$ and evanescent modes
with $k^{\prime\prime}<0$ are referred to as \emph{backward} waves. Since the
backward Bloch waves are not excited in the setting of Fig. \ref{SISn}, they
play no role in further consideration.

Note that the assumption that the transmitted wave $\Psi_{T}\left(  z\right)
$ is a superposition of propagating and/or evanescent Bloch eigenmodes may not
apply if the frequency $\omega$ exactly coincides with one of the stationary
point frequencies (\ref{SP}). For example, at frequency $\omega_{0}$ of
stationary inflection point, there are no evanescent eigenmodes at all, and
the transmitted wave $\Psi_{T}\left(  z\right)  $ is a (non-Bloch) Floquet
eigenmode linearly growing with $z$ \cite{PRB03}. The term non-Bloch means
that the respective field distribution does not comply with the relation
(\ref{BF}). Similar situation occurs at frequency $\omega_{d}$ of degenerate
band edge. At the same time, at any general frequency, including the vicinity
of any stationary point (\ref{SP}), the transmitted wave $\Psi_{T}\left(
z\right)  $ is a superposition of two Bloch eigenmodes, each of which is
either propagating, or evanescent.

In all three cases (\ref{PsiT=pr+pr} -- \ref{PsiT=pr+ev}), the contribution of
a particular Bloch eigenmode to the transmitted wave $\Psi_{T}$ depends on the
polarization $\Psi_{I}$ of the incident wave. One can always choose some
special incident wave polarization, such that only one of the two forward
Bloch modes is excited and the transmitted wave $\Psi_{T}$ is a single Bloch
eigenmode. In the next subsection we will see that there is no frozen mode
regime in the case of a single mode excitation. This fact relates to the very
nature of the frozen mode regime.

Knowing the eigenmode composition of the transmitted wave we can give a
semi-qualitative description of what happens when the frequency $\omega$ of
the incident wave approaches one of the stationary points (\ref{SP}) in Fig.
\ref{DRSP3}. A consistent mathematical analysis of the asymptotic behavior in
the vicinity of different stationary points is rather complicated. The details
can be found in \cite{WRM06} and references therein.

\subsubsection{Regular photonic band edge}

We start with the simplest case of a regular photonic band edge. There are two
different possibilities in this case, but none of them is associated with the
frozen mode regime. The first one relates to the trivial case where none of
the layers of the periodic structure displays an in-plane anisotropy or
gyrotropy. As the result, all Bloch eigenmodes are doubly degenerate with
respect to polarization. A detailed description of this case can be found in
textbooks on optics. Slightly different scenario occurs if some of the layers
are anisotropic or gyrotropic and, as a result, the polarization degeneracy is
lifted. Just below the band edge frequency $\omega_{g}$ in Fig. \ref{DRSP3}%
(a), the transmitted field $\Psi_{T}\left(  z\right)  $ is a superposition
(\ref{PsiT=pr+ev}) of one propagating and one evanescent Bloch modes. Due to
the boundary condition (\ref{BC}), the amplitude of the transmitted wave at
$z=0$ is comparable to that of the incident wave. In the case of a generic
polarization of the incident light, the amplitudes of the propagating and
evanescent Bloch components at $z=0$ are also comparable to each other and to
the amplitude of the incident light%
\begin{equation}
\left\vert \Psi_{pr}\left(  0\right)  \right\vert \sim\left\vert \Psi
_{ev}\left(  0\right)  \right\vert \sim\left\vert \Psi_{I}\right\vert ,\text{
at }\ \omega\leq\omega_{g}. \label{pr = ev = I}%
\end{equation}
As the distance $z$ from the surface increases, the evanescent component
$\Psi_{ev}\left(  z\right)  $ decays rapidly, while the amplitude of the
propagating component remains constant. Eventually, at a certain distance from
the slab surface, the transmitted wave $\Psi_{T}\left(  z\right)  $ becomes
very close to the propagating mode%
\begin{equation}
\Psi_{T}\left(  z\right)  \approx\Psi_{pr}\left(  z\right)  ,\text{ at }z\gg
L,\ \omega\leq\omega_{g}. \label{PsiT=Psipr}%
\end{equation}

The evanescent component $\Psi_{ev}$ of the transmitted wave does not display
any singularity at the band edge frequency $\omega_{g}$. The propagating mode
$\Psi_{pr}$ does develop a singularity associated with vanishing group
velocity at $\omega\rightarrow\omega_{g}-0$, but its amplitude remains finite
and comparable to that of the incident wave. At $\omega>\omega_{g}$, this
propagating mode turns into another evanescent mode in (\ref{PsiT=ev+ev}). The
bottom line is that none of the Bloch components of the transmitted wave
develops a large amplitude in the vicinity of a regular photonic band edge.
There is no frozen mode regime in this case.

\subsubsection{Stationary inflection point}

\begin{figure}[tbph]
\scalebox{0.8}{\includegraphics[viewport=0 0 500 180,clip]{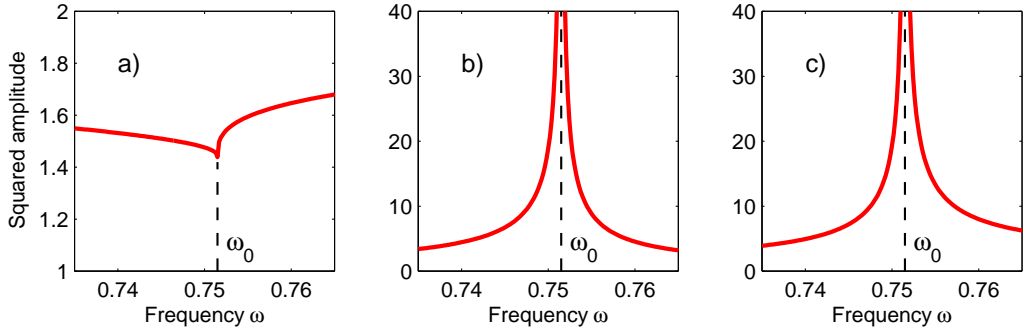}}
\caption{Frequency dependence of squared amplitude of the transmitted wave
(a), as well as its propagating (b) and evanescent (c) components at the
surface of semi-infinite photonic crystal at $z=0$. The stationary inflection
point frequency is $\omega_{0}$. Due to the destructive interference
(\ref{DI 3}) of the propagating and evanescent contributions, the amplitude
(a) of the resulting transmitted field remains small enough to satisfy the
boundary conditions. The amplitude of the incident wave is unity.}
\label{Am0_AAF}
\end{figure}

\begin{figure}[tbph]
\scalebox{0.8}{\includegraphics[viewport=0 0 500 180,clip]{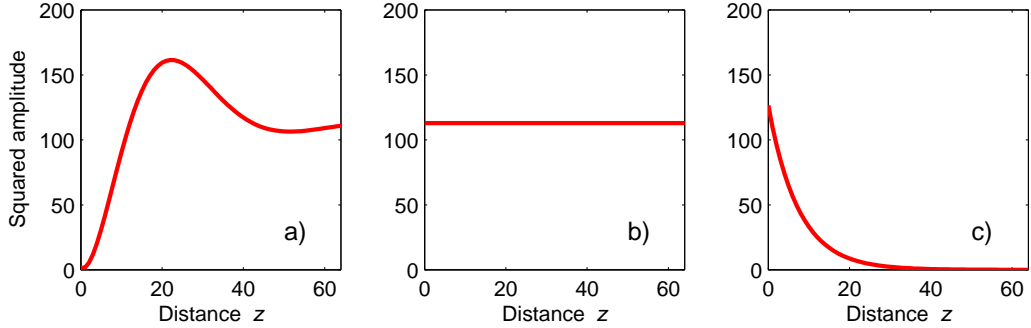}}
\caption{Destructive interference of the propagating and evanescent components
of the transmitted wave inside semi-infinite photonic crystal at $z>0$. The
frequency is close but not equal to that of the stationary inflection point.
(a) The squared amplitude $\left\vert \Psi_{T}\left(  z\right)  \right\vert
^{2}$ of the resulting transmitted field -- its amplitude at $z=0$ is small
enough to satisfy the boundary conditions; (b) the squared amplitude
$\left\vert \Psi_{pr}\left(  z\right)  \right\vert ^{2}$ of the propagating
contribution; (c) the squared amplitude $\left\vert \Psi_{ev}\left(  z\right)
\right\vert ^{2}$ of the evanescent contribution. The amplitude of the
incident wave is unity. The distance $z$ from the surface is expressed in
units of $L$.}
\label{Amz_AAF}
\end{figure}

A completely different situation develops in the vicinity of a stationary
inflection point (\ref{SIP DR}) of the dispersion relation. At $\omega
\approx\omega_{0}$, the transmitted wave $\Psi_{T}$ is a superposition
(\ref{PsiT=pr+ev}) of one propagating and one evanescent Bloch component. In
contrast to the case of a regular photonic band edge, in the vicinity of
$\omega_{0}$ both Bloch contributions to $\Psi_{T}$ develop strong
singularity. Specifically, as the frequency $\omega$ approaches $\omega_{0}$,
both contributions grow dramatically, while remaining nearly equal and
opposite in sign at the slab boundary \cite{PRB03}%
\begin{equation}
\Psi_{pr}\left(  0\right)  \approx-\Psi_{ev}\left(  0\right)  \propto
\left\vert \omega-\omega_{0}\right\vert ^{-1/3},\ \ \text{as }\omega
\rightarrow\omega_{0}. \label{DI 3}%
\end{equation}
Due to the destructive interference (\ref{DI 3}), the resulting field%
\[
\Psi_{T}\left(  0\right)  =\Psi_{pr}\left(  0\right)  +\Psi_{ev}\left(
0\right)
\]
at the photonic crystal surface is small enough to satisfy the boundary
condition (\ref{BC}), as illustrated in Figs. \ref{Am0_AAF} and \ref{Amz_AAF}.
As the distance $z$ from the slab boundary increases, the destructive
interference becomes less effective -- in part because the evanescent
contribution decays exponentially%
\begin{equation}
\Psi_{ev}\left(  z\right)  \approx\Psi_{ev}\left(  0\right)  \exp\left(
-zk^{\prime\prime}\right)  \label{Psi_ev SIP}%
\end{equation}
while the amplitude of the propagating contribution remains constant and very
large. Eventually, the transmitted wave $\Psi_{T}\left(  z\right)  $ reaches
its large saturation value corresponding to its propagating component
$\Psi_{pr}$, as illustrated in Fig. \ref{Amz_AAF}.

Note that the imaginary part $k^{\prime\prime}$ of the evanescent mode
wavenumber in (\ref{Psi_ev SIP}) also vanishes in the vicinity of stationary
inflection point%
\begin{equation}
k^{\prime\prime}\propto\left\vert \omega-\omega_{0}\right\vert ^{1/3}\text{,
as }\omega\rightarrow\omega_{0}, \label{Imk SIP}%
\end{equation}
reducing the rate of decay of the evanescent contribution (\ref{Psi_ev SIP}).
As a consequence, the resulting amplitude $\Psi_{T}\left(  z\right)  $ of the
transmitted wave reaches its large saturation value $\Psi_{pr}$ only at a
certain distance $Z$ from the surface. This characteristic distance increases
as the frequency approaches its critical value $\omega_{0}$%
\begin{equation}
Z\propto1/k^{\prime\prime}\propto\left\vert \omega-\omega_{0}\right\vert
^{-1/3}. \label{Z SIP}%
\end{equation}

If the frequency of the incident wave is exactly equal to the frozen mode
frequency $\omega_{0}$, the transmitted wave $\Psi_{T}\left(  z\right)  $ does
not reduce to the sum (\ref{PsiT=pr+ev}) of propagating and evanescent
contributions, because at $\omega=\omega_{0}$ there is no evanescent solutions
to the Maxwell equations. Instead, $\Psi_{T}\left(  z\right)  $ corresponds to
a non-Bloch Floquet eigenmode diverging linearly with $z$ \cite{PRB03}
\begin{equation}
\Psi_{T}\left(  z\right)  -\Psi_{T}\left(  0\right)  \propto z\Psi
_{0},\;\ \text{at}\;\omega=\omega_{0}.\label{FL SIP}%
\end{equation}%

\subsubsection{Degenerate band edge}

While the situation with a regular photonic band edge (\ref{RBE DR}) appears
trivial, the case of a degenerate band edge (\ref{DBE DR}) proves to be quite
different. Below the degenerate band edge frequency $\omega_{d}$, the
transmitted field is a superposition (\ref{PsiT=pr+ev}) of one propagating and
one evanescent components, while above $\omega_{d}$, the transmitted wave is a
combination (\ref{PsiT=ev+ev}) of two evanescent components. In this respect,
the regular and degenerate band edges are similar to each other. A crucial
difference, though, is that in the vicinity of a degenerate band edge, both
Bloch contributions to the transmitted wave diverge as $\omega$ approaches
$\omega_{d}$, which constitutes the frozen mode regime.

\begin{figure}[tbph]
\scalebox{0.8}{\includegraphics[viewport=0 0 500 180,clip]{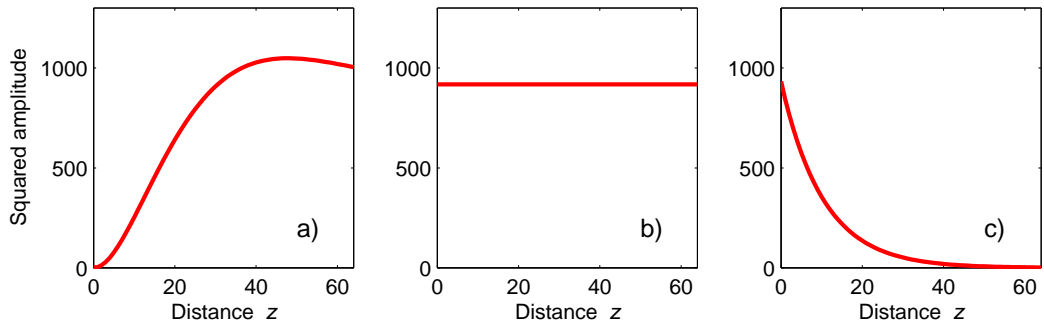}}
\caption{Smoothed filed distribution in the transmitted wave (a), along with
its propagating (b) and evanescent (c) components at frequency slightly below
that of the degenerate band edge in Fig. \ref{DRSP3}(c). The amplitude of the
incident wave is unity. Similar graphs related to the stationary inflection
point in Fig. \ref{DRSP3}(b) are shown in Fig.\ref{Am0_AAF}}
\label{Amn_B3}
\end{figure}

\begin{figure}[tbph]
\scalebox{0.8}{\includegraphics[viewport=0 0 500 400,clip]{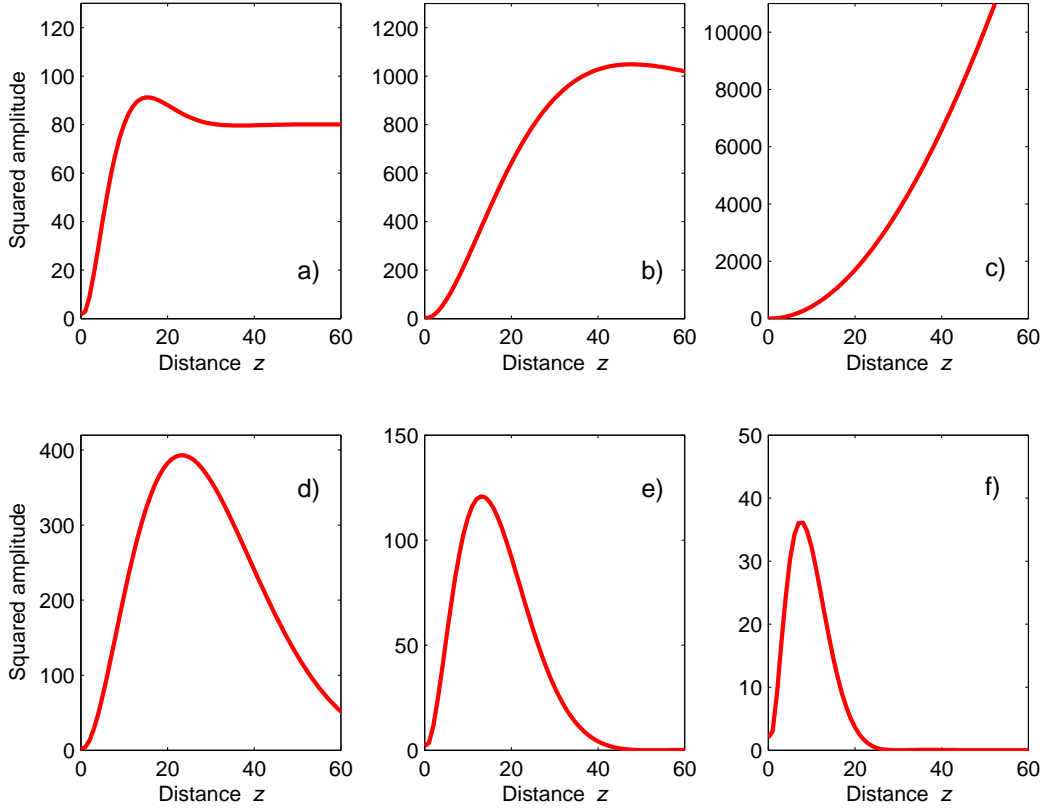}}
\caption{Smoothed profile of the frozen mode at six different frequencies in
the vicinity of degenerate band edge: (a) $\omega=\omega_{d}-10^{-4}c/L$, (b)
$\omega=\omega_{d}-10^{-6}c/L$, (c) $\omega=\omega_{d}$, (d) $\omega
=\omega_{d}+10^{-6}c/L$, (e) $\omega=\omega_{d}+10^{-5}c/L$, (f)
$\omega=\omega_{d}+10^{-4}c/L$. In the transmission band at $\omega<\omega
_{d}$, the asymptotic field value is very large and diverges as $\omega
\rightarrow\omega_{d}$. By contrast, in the band gap at $\omega>\omega_{d}$,
the asymptotic field value is zero. In all cases, the amplitude of the
incident wave is unity.}
\label{Amn6w}
\end{figure}

Let us start with the transmission band at $\omega<\omega_{d}$. As the
frequency $\omega$ approaches $\omega_{d}-0$, both Bloch contributions in
(\ref{PsiT=pr+ev}) grow sharply, while remaining nearly equal and opposite in
sign at the surface%
\begin{equation}
\Psi_{pr}\left(  0\right)  \approx-\Psi_{ev}\left(  0\right)  \propto
\left\vert \omega_{d}-\omega\right\vert ^{-1/4},\ \ \text{as }\omega
\rightarrow\omega_{d}-0. \label{DI 4b}%
\end{equation}
The destructive interference (\ref{DI 4b}) ensures that the boundary condition
(\ref{BC}) can be satisfied, while both Bloch contributions to $\Psi
_{T}\left(  z\right)  $ diverge. As the distance $z$ from the slab boundary
increases, the evanescent component $\Psi_{ev}\left(  z\right)  $ dies out%
\begin{equation}
\Psi_{ev}\left(  z\right)  \approx\Psi_{ev}\left(  0\right)  \exp\left(
-zk^{\prime\prime}\right)  \label{Psi_ev DBE}%
\end{equation}
while the propagating component $\Psi_{pr}\left(  z\right)  $ remains very
large. Eventually, as the distance $z$ further increases, the transmitted wave
$\Psi_{T}\left(  z\right)  $ reaches its large saturation value corresponding
to its propagating component $\Psi_{pr}\left(  z\right)  $, as illustrated in
Fig. \ref{Amn_B3}. Note that the imaginary part $k^{\prime\prime}$ of the
evanescent mode wavenumber also vanishes in the vicinity of degenerate band
edge%
\begin{equation}
k^{\prime\prime}\propto\left\vert \omega-\omega_{d}\right\vert ^{1/4}\text{,
as }\omega\rightarrow\omega_{d}, \label{Imk DBE}%
\end{equation}
reducing the rate of decay of the evanescent contribution (\ref{Psi_ev DBE}).
As a consequence, the resulting amplitude $\Psi_{T}\left(  z\right)  $ of the
transmitted wave reaches its large saturation value $\Psi_{pr}$ only at a
certain distance $Z$ from the surface. This characteristic distance increases
as the frequency approaches its critical value $\omega_{d}$%
\begin{equation}
Z\propto1/k^{\prime\prime}\propto\left\vert \omega-\omega_{d}\right\vert
^{-1/4}, \label{Z DBE}%
\end{equation}
as illustrated in Fig. \ref{Amn6w}(a, b). If the frequency $\omega$ of the
incident wave is exactly equal to $\omega_{d}$, the transmitted wave $\Psi
_{T}\left(  z\right)  $ does not reduce to the sum of two Bloch contributions.
Instead, it corresponds to a non-Bloch Floquet eigenmode linearly diverging
with $z$
\begin{equation}
\Psi_{T}\left(  z\right)  -\Psi_{T}\left(  0\right)  \propto z\Psi
_{d},\;\ \text{at}\;\omega=\omega_{d}. \label{FL DBE}%
\end{equation}
This situation is illustrated in Fig. \ref{Amn6w}(c).

The above behavior appears to be very similar to that of the frozen mode
regime at a stationary inflection point, shown in Fig. \ref{Amz_AAF}. Yet,
there is a crucial difference between the two cases. Indeed, according to
(\ref{ST(DBE)}), in the immediate proximity of a degenerate band edge, the
Pointing vector $S_{T}$ of the transmitted wave $\Psi_{T}$ is infinitesimal,
in spite of the diverging wave amplitude (\ref{DI 4b}). In other words,
although the energy density $W_{T}\propto\left\vert \Psi_{T}\right\vert ^{2}$
of the frozen mode diverges as $\omega\rightarrow\omega_{d}-0$, it does not
grow fast enough to offset the vanishing group velocity (\ref{u(DBE)}). As a
consequence, the photonic crystal becomes totally reflective at $\omega
=\omega_{d}$. Of course, the total reflectivity persists at $\omega>\omega
_{d}$, where there is no propagating modes at all. By contrast, in the case
(\ref{FL SIP}) of a stationary inflection point, the respective Pointing
vector $S_{T}$ is finite and can be even close to that of the incident wave.
The latter implies low reflectivity and nearly total conversion of the
incident wave energy into the frozen mode.

\begin{figure}[tbph]
\scalebox{0.8}{\includegraphics[viewport=0 0 500 180,clip]{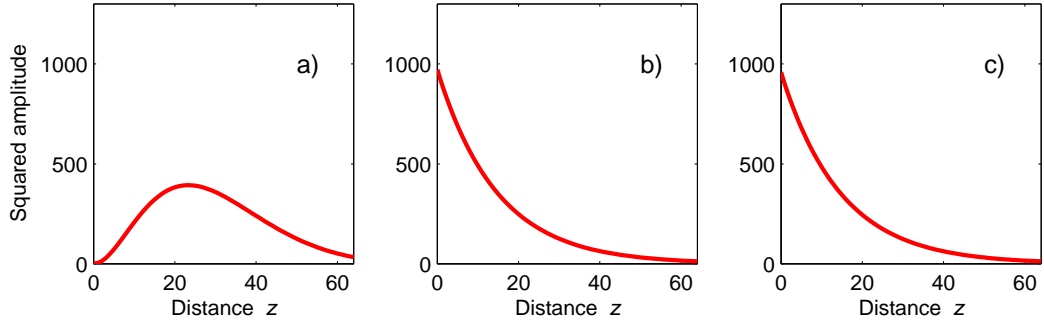}}
\caption{Smoothed filed distribution in the transmitted wave (a), along with
its two evanescent components (b) and (c), at a band gap frequency just above
the degenerate band edge value $\omega_{d}$ in Fig. \ref{DRSP3}(c). Compare
these graphs with those in Fig. \ref{Amn_B3}, where the frequency lies in the
transmission band below $\omega_{d}$.}
\label{Amn_G3}
\end{figure}

Interesting situation occurs when we approach the degenerate band edge
frequency from the band gap. In such a case, the transmitted field $\Psi
_{T}\left(  z\right)  $ is a superposition (\ref{PsiT=ev+ev}) of two
evanescent components. As the frequency $\omega$ approaches $\omega_{d}$. Both
evanescent contributions grow sharply, while remaining nearly equal and
opposite in sign at the photonic crystal boundary%
\begin{equation}
\Psi_{ev1}\left(  0\right)  \approx-\Psi_{ev2}\left(  0\right)  \propto
\left\vert \omega_{d}-\omega\right\vert ^{-1/4},\ \ \text{as }\omega
\rightarrow\omega_{d}+0.\label{DI 4g}%
\end{equation}
Again, the destructive interference (\ref{DI 4g}) ensures that the boundary
condition (\ref{BC}) is satisfied, while both evanescent contributions to
$\Psi_{T}\left(  z\right)  $ diverge in accordance with (\ref{DI 4g}). As the
distance $z$ from the slab boundary increases, the destructive interference of
these two evanescent components is lifted and the resulting field amplitude
increases sharply, as shown in Fig. \ref{Amn_G3}. But eventually, as the
distance $z$ further increases, the transmitted wave $\Psi_{T}\left(
z\right)  $ completely decays, because both Bloch contributions to $\Psi
_{T}\left(  z\right)  $ are evanescent. The latter constitutes the major
difference between the frozen mode regime above and below the DBE frequency
$\omega_{d}$. The rate of the amplitude decay, as well as the position of the
maximum of the transmitted wave amplitude in Fig. \ref{Amn_G3}(a), are
determined by the characteristic distance $Z$ in (\ref{Z DBE}).%

\subsubsection{Physical reason for the growing wave amplitude}

The wave $\Psi_{T}\left(  z\right)  $ transmitted to the semi-infinite
periodic layered medium is a superposition of two forward Bloch modes
$\Psi_{1}\left(  z\right)  $ and $\Psi_{2}\left(  z\right)  $%
\begin{equation}
\Psi_{T}\left(  z\right)  =\Psi_{1}\left(  z\right)  +\Psi_{2}\left(
z\right)  . \label{Psi1+Psi2}%
\end{equation}
The two eigenmodes in (\ref{Psi1+Psi2}) can be a propagating and an
evanescent, as in the case (\ref{PsiT=pr+ev}), or they can be both evanescent,
as in the case (\ref{PsiT=ev+ev}). The physical reason for the sharp increase
in the transmitted wave amplitude at the frozen mode regime is that the two
Bloch eigenmodes contributing to $\Psi_{T}\left(  z\right)  $ become nearly
indistinguishable from each other, as the frequency approaches its critical
value%
\begin{equation}
\Psi_{1}\left(  z\right)  \approx\alpha\Psi_{2}\left(  z\right)  ,\ \ \text{as
}\omega\rightarrow\omega_{s}, \label{Psi - Psi}%
\end{equation}
where $\alpha$ is a scalar, and $\omega_{s}$ is the frozen mode frequency
($\omega_{0}$ or $\omega_{d}$).

Let us show under what circumstances the property (\ref{Psi - Psi}) can lead
to the frozen mode regime. The sum (\ref{Psi1+Psi2}) of two nearly parallel
column vectors $\Psi_{1}$ and $\Psi_{2}$ must match the boundary conditions
(\ref{BC}) with the incident and reflected waves. If the incident wave
polarization is general, then the nearly parallel Bloch components $\Psi_{1}$
and $\Psi_{2}$ must be very large and nearly equal and opposite%
\begin{equation}
\Psi_{1}\left(  0\right)  \approx-\Psi_{2}\left(  0\right)  ,\ \ \left\vert
\Psi_{1}\left(  0\right)  \right\vert \approx\left\vert \Psi_{2}\left(
0\right)  \right\vert \gg\left\vert \Psi_{I}\right\vert ,\label{Psi1= -Psi2}%
\end{equation}
in order to satisfy the boundary conditions (\ref{BC}). Indeed, since the
incident field polarization is general, we have no reason to expect that the
column vector $\Psi\left(  0\right)  $ at the surface is nearly parallel to
$\Psi_{1}\left(  0\right)  $ and $\Psi_{2}\left(  0\right)  $. But on the
other hand, the boundary conditions say that%
\begin{equation}
\Psi\left(  0\right)  =\Psi_{1}\left(  0\right)  +\Psi_{2}\left(  0\right)
\label{Psi(0)}%
\end{equation}
Obviously, the only situation where the sum (\ref{Psi(0)}) of two nearly
parallel vectors can be not nearly parallel to either of them is the one
described in (\ref{Psi1= -Psi2}).

There is one exception, though. Recall that in the vicinity of the frozen mode 
frequency, the two Bloch components $\Psi_{1}$ and $\Psi_{2}$ of the transmitted 
wave are nearly parallel to each other (see formula (\ref{Psi - Psi})). For this
reason, if the polarization of the incident wave is such that
$\Psi\left(  0\right)$ in (\ref{Psi(0)}) is nearly parallel
to any of the Bloch eigenmodes $\Psi_{1}\left(  0\right)  $ and $\Psi
_{2}\left(  0\right)  $, it is also nearly parallel to both
of them. In this, and only this case, the amplitude of the transmitted wave
$\Psi_{T}\left(  z\right)  $ will be comparable to that of the incident wave.
There is no frozen mode regime for this narrow range of the incident wave
polarization. A particular case of the above situation is a regime of a single
mode excitation, where only one of the two Bloch components $\Psi_{1}$ or
$\Psi_{2}$ in (\ref{Psi1+Psi2}) contributes to the transmitted wave.

Finally, let us reiterate that in the limiting cases of $\omega=\omega_{0}$ or
$\omega=\omega_{d}$, the transmitted wave $\Psi_{T}\left(  z\right)  $
corresponds to a non-Bloch Floquet eigenmode (\ref{FL SIP}) or (\ref{FL DBE}),
respectively. Either of them linearly diverges with $z$. Again, the only
exception is when the incident wave has the unique polarization, at which the
transmitted wave $\Psi_{T}\left(  z\right)  $ is a propagating Bloch eigenmode
with zero group velocity and a limited amplitude, comparable to that of the
incident wave.

\section{Oblique propagation. Abnormal grazing modes}

\begin{figure}[tbph]
\scalebox{0.8}{\includegraphics[viewport=-100 0 500 200,clip]{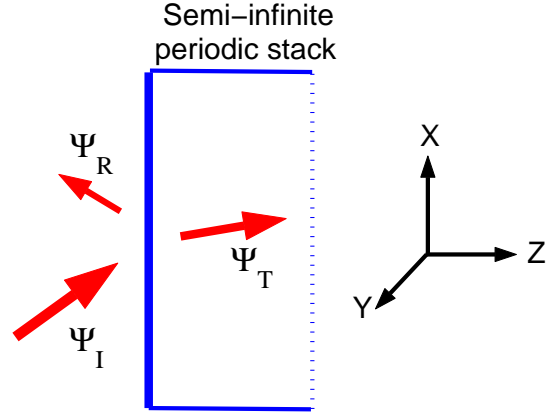}}
\caption{Scattering problem for a plane wave obliquely incident on a
semi-infinite periodic layered medium. The arrows schematically shows the
Pointing vectors of the incident, reflected and transmitted waves. The energy
flux $S_{I}$ of the incident plane wave is unity.}
\label{SISob}
\end{figure}

A\ phenomenon similar to the frozen mode regime can also occur in the case of
oblique wave propagation, where the incident, reflected and transmitted waves
are all propagate at an angle to the $z$ axis, as shown in Fig. \ref{SISob}.
Consider the situation where the normal component $u_{z}$ of the group
velocity of the transmitted propagating wave vanishes, while the tangential
component $\vec{u}_{\perp}$ remains finite.%
\begin{equation}
u_{z}=\frac{\partial\omega}{\partial k_{z}}=0,\ \ \vec{u}_{\perp}%
=\frac{\partial\omega}{\partial\vec{k}_{\perp}}\neq0\text{, at }\omega
=\omega_{s}=\omega\left(  \vec{k}_{s}\right)  \label{ASP}%
\end{equation}
This is exactly what happens in the vicinity of the well-known phenomenon of
total internal reflection \cite{LLEM}. Similar phenomenon can be found in any
photonic crystal at frequency corresponding to the transmission band edge for
a particular direction of incidence. Remarkably, the total reflection of the
incident wave is not the only possible outcome. Another alternative is that
the transmitted wave forms an abnormal grazing mode with dramatically enhanced
amplitude and tangential group velocity. The photonic crystal reflectivity at
the critical point (\ref{ASP}) can be even low -- quite the opposite to what
happens at total internal reflection. Low reflectivity implies that in spite
of the vanishing normal component $u_{z}$ of the transmitted wave group
velocity, almost all the energy of the incident wave is converted into an
abnormal grazing mode with drastically enhanced amplitude. The profile of such
a grazing mode, i.e., the field dependence on the distance $z$ from the
surface, appears to be very similar to that of the frozen mode at normal
incidence shown in Figs. \ref{Amz_AAF}, \ref{Amn_B3}, \ref{Amn_G3}, and
\ref{Amn6w}. The only difference is that the tangential component of the group
velocity at an oblique version of the frozen mode regime can be large and
comparable to the speed of light in vacuum. A significant advantage of the
oblique version of the frozen mode regime is that it can occur in much simpler
periodic structures, compared to those supporting the frozen mode regime at
normal incidence. Examples of periodic layered arrays supporting only the
oblique version of the frozen mode regime are shown in Figs. \ref{StackAB} and
\ref{StackABc}. These structures are too simple to support any kind of frozen
mode regime at normal incidence -- they have only two different layers per
unit cell, of which only one layer is anisotropic. But at oblique incidence,
these relatively simple periodic arrays can support the frozen mode regime.
The presence of at least one anisotropic layer in a unit cell $L$ is still required.

Further in this section we outline some basic properties of the frozen mode
regime at oblique incidence.%

\begin{figure}[tbph]
\scalebox{0.8}{\includegraphics[viewport=-100 0 500 200,clip]{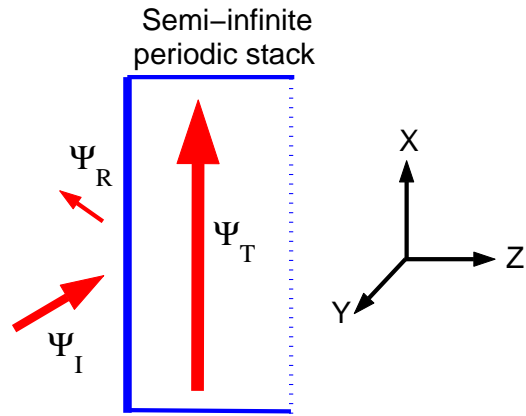}}
\caption{The case of oblique incidence, where the transmitted wave is a
grazing mode with very large amplitude and purely tangential energy flux.}
\label{SSIt}
\end{figure}

\subsection{Axial dispersion relation: basic definitions}

Consider a plane monochromatic wave obliquely incident on a periodic
semi-infinite stack, as shown in Fig. \ref{SISob}. Due to the boundary
conditions (\ref{BC}), the incident, reflected and transmitted waves should be
assigned the same pair of tangential components $k_{x},k_{y}$ of the
respective wave vectors%
\begin{equation}
\left(  \vec{k}_{I}\right)  _{x}=\left(  \vec{k}_{R}\right)  _{x}=\left(
\vec{k}_{T}\right)  _{x},\ \ \left(  \vec{k}_{I}\right)  _{y}=\left(  \vec
{k}_{R}\right)  _{y}=\left(  \vec{k}_{T}\right)  _{y}, \label{BC kx ky}%
\end{equation}
while the axial (normal) components $k_{z}$ are all different. For the
incident and reflected waves we have simply%
\begin{equation}
\left(  \vec{k}_{I}\right)  _{z}=-\left(  \vec{k}_{R}\right)  _{z}%
=\sqrt{\omega^{2}c^{2}-k_{x}^{2}-k_{y}^{2}}. \label{kz}%
\end{equation}

Let us turn to the transmitted wave. The transmitted wave is usually a
composition of two Bloch eigenmodes with the same $\vec{k}_{\perp}=\left(
k_{x},k_{y}\right)  $ from (\ref{BC kx ky}), but different $k_{z}$. For given
$\vec{k}_{\perp}$ and $\omega$, the value of $k_{z}$ is obtained by solving
the time-harmonic Maxwell equations in the periodic medium. The so found
correspondence between the wavenumber $k_{z}$ and the frequency $\omega$ at
fixed $\vec{k}_{\perp}$ is referred to as the \emph{axial} or \emph{normal}
dispersion relation. Real $k_{z}$ correspond to propagating (traveling) Bloch
modes, while complex $k_{z}$ correspond to evanescent modes, decaying with the
distance $z$ from from the surface. Unlike $k_{x}$ and $k_{y}$, the Bloch
wavenumber $k_{z}$ is defined up to a multiple of $2\pi/L$.

Similarly to the case of normal propagation, the expression (\ref{ASP})
defines stationary points of the axial dispersion relation. The definition
(\ref{ASP}) is a generalization of (\ref{SP}) to the case of oblique
propagation. Different kinds of axial stationary points are defined as follows.

\begin{itemize}
\item[-] A regular band edge of axial dispersion relation%
\begin{equation}
\frac{\partial\omega}{\partial k_{z}}=0,\ \frac{\partial^{2}\omega}{\partial
k_{z}^{2}}\neq0 \label{A RBE}%
\end{equation}

\item[-] A stationary inflection point of axial dispersion relation%
\begin{equation}
\frac{\partial\omega}{\partial k_{z}}=0,\ \frac{\partial^{2}\omega}{\partial
k_{z}^{2}}=0,\ \frac{\partial^{3}\omega}{\partial k_{z}^{3}}\neq0
\label{A SIP}%
\end{equation}

\item[-] A degenerate band edge of axial dispersion relation%
\begin{equation}
\frac{\partial\omega}{\partial k_{z}}=0,\ \frac{\partial^{2}\omega}{\partial
k_{z}^{2}}=0,\ \frac{\partial^{3}\omega}{\partial k_{z}^{3}}=0,\ \frac
{\partial^{3}\omega}{\partial k_{z}^{3}}\neq0 \label{A DBE}%
\end{equation}

\end{itemize}

The above definitions are analogous to those in (\ref{RBE DR}), (\ref{SIP DR}%
), and (\ref{DBE DR}), related to the case of normal propagation. We still can
refer to the $k-\omega\ $diagrams in Fig. \ref{DRSP3}, where the quantity $k$
is now understood as the normal component $k_{z}$ of the Bloch wavenumber at
fixed $\vec{k}_{\perp}$.

In order to further establish a close analogy between the cases of normal and
oblique propagation, let us introduce some standard procedure commonly used in
electrodynamics of layered media. In the case of oblique propagation, the
time-harmonic electric and magnetic fields $\vec{E}\left(  x,y,z\right)  $ and
$\vec{H}\left(  x,y,z\right)  $ depend on all three Cartesian coordinates,
both inside (at $z>0$) and outside (at $z<0$) the periodic medium. Given the
boundary conditions (\ref{BC kx ky}), the field dependence on the transverse
coordinates $x$ and $y$ can be accounted for by the following substitution in
the time-harmonic Maxwell equations%
\begin{equation}
\vec{E}\left(  \vec{r}\right)  =e^{i\left(  k_{x}x+k_{y}y\right)
}\mathcal{\vec{E}}\left(  z\right)  ,\ \vec{H}\left(  \vec{r}\right)
=e^{i\left(  k_{x}x+k_{y}y\right)  }\mathcal{\vec{H}}\left(  z\right)  .
\label{LEM}%
\end{equation}
The next step is a separation of the transverse components of $\mathcal{\vec
{E}}\left(  z\right)  $ and $\mathcal{\vec{H}}\left(  z\right)  $ into a close
system of four linear differential equations%
\begin{equation}
\partial_{z}\Psi\left(  z\right)  =i\frac{\omega}{c}M\left(  z\right)
\Psi\left(  z\right)  ,\; \label{ME4}%
\end{equation}
where $\Psi\left(  z\right)  $ is a vector-column%
\begin{equation}
\Psi\left(  z\right)  =\left[
\begin{array}
[c]{c}%
\mathcal{E}_{x}\left(  z\right) \\
\mathcal{E}_{y}\left(  z\right) \\
\mathcal{E}_{x}\left(  z\right) \\
\mathcal{E}_{y}\left(  z\right)
\end{array}
\right]  \label{Psi obl}%
\end{equation}
similar to that in (\ref{Psi}). An explicit expression for the matrix
$M\left(  z\right)  $ can be found, for example, in Ref. \cite{WRM06}, along
with the detailed discussion of its analytical properties.

In the case of oblique incidence, the $4\times4$ matrix $M\left(  z\right)  $
in the time-harmonic Maxwell equations (\ref{ME4}) depends not only on
physical parameters of the periodic structure, but also on the tangential
components $\vec{k}_{\perp}=\left(  k_{x},k_{y}\right)  $ of the wave vector.
In the particular case of normal propagation, the symmetry of the $4\times4$
matrix $M\left(  z\right)  $ in (\ref{ME4}) increases so that it may become
incompatible with the existence of a stationary inflection point
(\ref{SIP DR}) and/or a degenerate band edge (\ref{DBE DR}). This is why the
axial frozen mode regime at oblique incidence can occur in some periodic
structures which are too simple to support the normal frozen mode regime.
Examples such structures are presented in Figs. \ref{StackAB} and
\ref{StackABc}.

All basic features of axially frozen mode regime at oblique incidence are
virtually the same as in the case of normal propagation. In particular, all
the formulae of the previous section describing the structure and composition
of the transmitted field $\Psi_{T}\left(  z\right)  $ remain unchanged. This
close similarity holds both for the frozen mode regime at a stationary
inflection point (\ref{A SIP}) and at a degenerate band edge (\ref{A DBE}). In
either case, Figs. \ref{Amz_AAF} through \ref{Amn6w} give an adequate idea of
the frozen mode structure. Still, there is one essential difference. Namely,
in the case of axially frozen mode we have to remember that the tangential
component of the group velocity is not zero, even if the normal component
(\ref{ASP}) vanishes. This means that the axially frozen mode is in fact an
abnormal grazing mode with purely tangential energy flux, greatly enhanced
amplitude, and very unusual profile. In the next subsection we elaborate on
this point.

\subsection{Axially frozen mode as an abnormal grazing wave}

\subsubsection{Stationary inflection point of axial dispersion relation}

Let us start with the case (\ref{SIP DR}) of stationary inflection point. In
the vicinity of $\omega_{0}$, the transmitted wave $\Psi_{T}\left(  z\right)
$ is a superposition (\ref{PsiT=pr+ev}) of propagating and evanescent Bloch
eigenmodes. According to (\ref{DI 3}), both contributions to $\Psi_{T}\left(
z\right)  $ have huge and nearly equal and opposite values near the photonic
crystal boundary, so that $\Psi_{T}\left(  0\right)  $ is small enough to
satisfy the boundary condition (\ref{BC}). As the distance $z$ from the slab
boundary increases, the evanescent component $\Psi_{ev}\left(  z\right)  $
decays exponentially, while the amplitude of the propagating component
$\Psi_{pr}\left(  z\right)  $ remains constant and very large. As the result,
the field amplitude reaches its huge saturation value at a certain distance
(\ref{Z SIP}) from the slab boundary, as illustrated in Fig. \ref{Amz_AAF}.

Let us now introduce a unit vector $\vec{n}$ parallel to the direction of
incident wave propagation%
\begin{equation}
n_{x}=k_{x}c/\omega,\ \ n_{y}=k_{y}c/\omega,\ \ n_{z}=\sqrt{1-\left(
n_{x}^{2}+n_{y}^{2}\right)  },\label{n(k)}%
\end{equation}
and let $\vec{n}_{0}$ and $\omega_{0}$ be the direction of incidence and the
frequency corresponding to a stationary inflection point (\ref{A SIP}) of the
axial dispersion relation. When the frequency $\omega$ tends to its critical
value $\omega_{0}$ at fixed direction of incidence $\vec{n}_{0}$, the
saturation value $\Psi_{pr}$ of the transmitted wave amplitude diverges as
$\left\vert \omega-\omega_{0}\right\vert ^{-1/3},$ according to (\ref{DI 3}).
This behavior is illustrated in Fig. \ref{Am0_AAF}. Conversely, when the
direction of incidence $\vec{n}$ tends to its critical value $\vec{n}_{0}$ at
fixed frequency $\omega_{0}$, the respective saturation value $\Psi_{pr}$ of
the transmitted wave amplitude diverges as $\left\vert \vec{n}-\vec{n}%
_{0}\right\vert ^{-1/3}$.

\begin{figure}[tbph]
\scalebox{0.6}{\includegraphics[viewport=-50 0 500 400,clip]{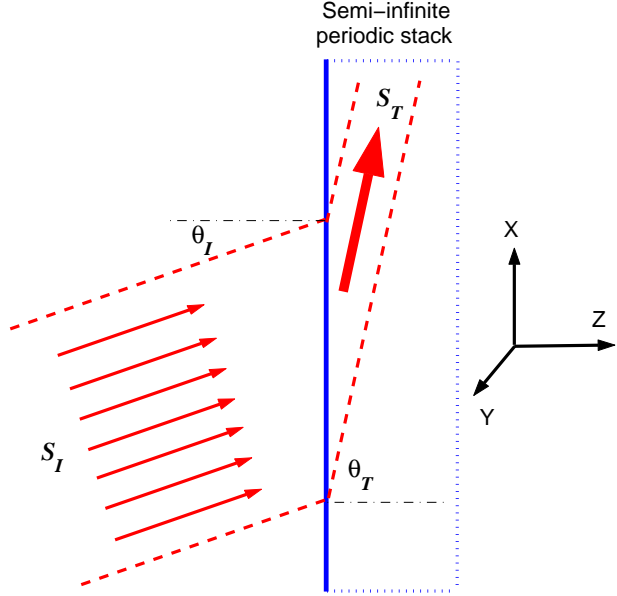}}
\caption{Incident and transmitted (refracted) waves in the vicinity of the%
 axial frozen mode regime
regime. The reflected wave is not shown. $\theta_{I}$ and $\theta_{T}$ are the
incidence and refraction angles, $S_{I}$ and $S_{T}$ are the energy density
fluxes of the incident and transmitted waves. Both the energy density and the
energy density flux in the transmitted wave are much larger than the
respective values in the incident wave. However, the total power transmitted
by the refracted wave is smaller by factor $\tau$, due to much smaller
cross-section area of the nearly grazing transmitted wave.}
\label{Beam}
\end{figure}

Consider now the frozen mode regime at oblique incidence in terms of
refraction of the incident plane wave at the photonic crystal boundary. In
Fig. \ref{Beam} we show a wide monochromatic beam of frequency $\omega$
incident on the surface of the semi-infinite photonic crystal. The refraction
angle $\theta_{T}$ is determined by the ratio of the normal and tangential
components of the group velocity%
\begin{equation}
\frac{\pi}{2}-\theta_{T}=\arctan\frac{u_{z}}{u_{\perp}}. \label{theta_T}%
\end{equation}
The direction of incidence $\vec{n}_{0}\parallel\vec{S}_{I}$ in Fig.
\ref{Beam} is chosen so that the condition (\ref{A SIP}) of the frozen mode
regime is satisfied at $\omega=\omega_{0}$. As frequency $\omega$ tends to
$\omega_{0}$ from either direction, the normal component $u_{z}$ of the
transmitted wave group velocity vanishes, while the tangential component
$\vec{u}_{\perp}$ remains finite
\begin{equation}
u_{z}\thicksim\left\vert \omega-\omega_{0}\right\vert ^{2/3}\rightarrow
0,\;\vec{u}_{\perp}\rightarrow\vec{u}_{0}\;\text{\ \ as \ }\omega
\rightarrow\omega_{0}. \label{u_z, u_n}%
\end{equation}
This relation together with (\ref{theta_T}) yield%
\begin{equation}
\frac{\pi}{2}-\theta_{T}\propto\left\vert \omega-\omega_{0}\right\vert
^{2/3}\rightarrow0\;\text{\ \ as \ }\omega\rightarrow\omega_{0}.
\label{theta 1}%
\end{equation}

Evidently, in the vicinity of the frozen mode frequency, the transmitted
(refracted) propagating mode can be viewed as a grazing mode, because its
group velocity becomes parallel to the surface.

The most important and unique feature of this abnormal grazing mode directly
relates to the fact that the transmittance of the photonic crystal remains
finite even at $\omega=\omega_{0}$. Indeed, let $A_{I}$ and $A_{T}$ be the
cross-section areas of the incident and transmitted (refracted) beams in Fig.
\ref{Beam}, respectively. Obliviously,%
\begin{equation}
\frac{A_{T}}{A_{I}}=\frac{\cos\theta_{T}}{\cos\theta_{I}} \label{theta 2}%
\end{equation}
Let us also introduce the quantities
\begin{equation}
U_{I}=A_{I}S_{I},\;U_{T}=A_{T}S_{T}, \label{theta 3}%
\end{equation}
where $S_{I}$ and $S_{T}$ are the energy fluxes of the incident and
transmitted waves. $U_{I}$ and $U_{T}$ are the total power transmitted by the
incident and refracted beams, respectively. The expressions (\ref{theta 2})
and (\ref{theta 3}) imply that%
\begin{equation}
\frac{U_{T}}{U_{I}}=\frac{S_{T}\cos\theta_{T}}{S_{I}\cos\theta_{I}}%
=\frac{\left(  S_{T}\right)  _{z}}{\left(  S_{I}\right)  _{z}}=\tau
\label{theta 4}%
\end{equation}
which is nothing more than a manifestation of the energy conservation.
Equation (\ref{theta 4}) together with (\ref{theta 1}) yield%
\begin{equation}
S_{T}=\tau S_{I}\frac{\cos\theta_{I}}{\cos\theta_{T}}\propto\left\vert
\omega-\omega_{0}\right\vert ^{-2/3}\rightarrow\infty\;\text{\ \ as \ }%
\omega\rightarrow\omega_{0}. \label{theta 5}%
\end{equation}
where we have taken into account that $\tau\cos\theta_{I}$ is of the order of
unity, as \ $\omega\rightarrow\omega_{0}$.

The expressions (\ref{theta 1} -- \ref{theta 5}) show that in the vicinity of
the frozen mode regime, the transmitted beam is a grazing mode with very large
intensity $S_{T}$ and sharply reduce cross-section area $A_{T}$, compared to
those of the incident beam. By contrast, in the vicinity of the photonic band
edge the transmittance $\tau$ of the semi-infinite slab vanishes along with
the normal component of the energy flux $\vec{S}_{T}$ of the transmitted
(refracted) wave.

The above qualitative analysis is only valid on the scales larger than the
characteristic length $Z$ in (\ref{Z SIP}). The value $Z$ defines the distance
from photonic crystal surface, where the evanescent contribution to the
transmitted wave is still significant. The latter restriction implies that the
width of both the incident and the refracted beams must be much larger than
$Z$. Otherwise, the transmitted wave cannot be treated as a beam, and the
expressions (\ref{theta 1} -- \ref{theta 5}) do not apply.%

\subsubsection{Degenerate band edge of axial dispersion relation}

Below the degenerate band edge frequency $\omega_{d}$ in (\ref{A DBE}), the
transmitted wave $\Psi_{T}\left(  z\right)  $ is a superposition
(\ref{PsiT=pr+ev}) of propagating and evanescent Bloch eigenmodes. According
to (\ref{DI 4b}), both contributions to $\Psi_{T}\left(  z\right)  $ have very
large and nearly equal and opposite values near the photonic crystal boundary
at $z=0$, so that $\Psi_{T}\left(  0\right)  $ is small enough to satisfy the
boundary condition (\ref{BC}). As the distance $z$ from the slab boundary
increases, the evanescent component $\Psi_{ev}\left(  z\right)  $ decays
exponentially, while the amplitude of the propagating component $\Psi
_{pr}\left(  z\right)  $ remains constant and very large. As the result, the
field amplitude reaches its huge saturation value at a certain distance from
the slab boundary, as illustrated in Fig. \ref{Amz_AAF}. At first sight, this
scenario is similar to the case of oblique frozen mode regime at stationary
inflection point of the axial dispersion relation. But there is an important
difference. As the frequency $\omega$ tends to its critical value $\omega_{d}%
$, the photonic crystal reflectivity approaches unity, implying total
reflection of the incident radiation. In this respect, the situation below an
axial degenerate band edge is reminiscent of the classical phenomenon of total
internal reflection. The difference, though, is that the transmitted wave
profile $\Psi_{T}\left(  z\right)  $ is the same as in the previously
discussed case of stationary inflection point of the axial dispersion
relation. It implies that just below a degenerate band edge frequency of axial
dispersion relation, the transmitted wave can be classified as an abnormal
grazing mode with large diverging amplitude and large tangential component of
the energy flux. The typical profile of such a grazing mode at $\omega
\lessapprox\omega_{d}$ is shown in Figs. \ref{Amn_B3}(a) and \ref{Amn6w}(a -- b).

If the incident wave frequency lies slightly above $\omega_{d}$ (in the band
gap), the transmitted wave $\Psi_{T}\left(  z\right)  $ is a superposition
(\ref{PsiT=ev+ev}) of two evanescent Bloch eigenmodes. The energy flux in the
abnormal grazing mode now is strictly parallel to the surface, regardless of
small variations in frequency and/or direction of propagation of the incident
wave. This is a direct consequence of the total reflectivity of photonic
crystals at band gap frequencies. The grazing mode profile in this case is
shown in Figs. \ref{Amn_G3}(a) and \ref{Amn6w}(d -- f). As the frequency
$\omega$ approaches its critical value $\omega_{d}$, the amplitude and the
thickness of the grazing mode diverge in accordance with (\ref{DI 4g}) and
(\ref{Z DBE}), respectively.

\subsubsection{Surface waves in the vicinity of a degenerate band edge of
axial dispersion relation}

So far in this section we have tacitly assumed that%
\begin{equation}
k_{x}^{2}+k_{y}^{2}<\omega^{2}c^{2}.\label{kt<wc}%
\end{equation}
The inequality (\ref{kt<wc}) implies that the $z$ component (\ref{kz}) of the
wave vector of the incident and reflected waves is real. This is a natural
assumption when considering the problem of a plane wave incident on a
semi-infinite photonic crystal.

Consider now the opposite case where%
\begin{equation}
k_{x}^{2}+k_{y}^{2}>\omega^{2}c^{2}.\label{kt>wc}%
\end{equation}
In this situation, there is no plane propagating waves in vacuum matching the
boundary conditions (\ref{BC kx ky}) at the surface. Still, if the frequency
$\omega$ lies above the band edge of the axial dispersion relation (in the
band gap), a superposition (\ref{PsiT=ev+ev}) of two forward evanescent
eigenmodes with $k_{z}^{\prime\prime}>0$ can produce a surface wave
exponentially decaying with the distance $z$ from the boundary in either
direction. As the frequency $\omega$ approaches the degenerate band edge value
$\omega_{d}$ for a given $\vec{k}_{\perp}$, the surface wave profile can
become similar to that of an abnormal grazing mode at $\omega\gtrapprox
\omega_{d}$, shown in Figs. \ref{Amn_G3}(a) and \ref{Amn6w}(d -- f). Although
formally, it is still a surface wave, its profile now is highly unusual for a
surface wave. Namely, the field amplitude in the periodic medium sharply
increases with the distance $z$ from the surface, reaches its maximum at a
distance $Z$ defined in (\ref{Z DBE}), and only after that it begins a slow
decay.%

\begin{figure}[tbph]
\scalebox{0.8}{\includegraphics[viewport=0 0 500 400,clip]{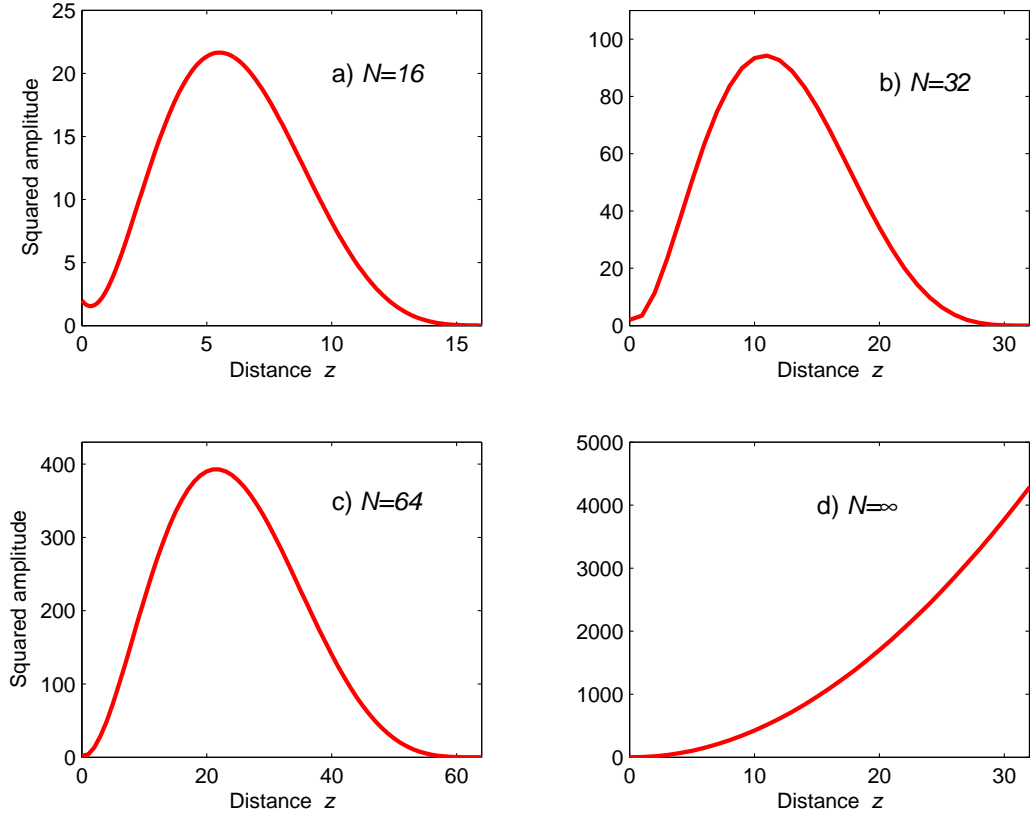}}
\caption{Smoothed profile of the frozen mode in periodic layered structures
composed of different number $N$ of unit cells $L$ in Fig. \ref{StackAAB}. The
frequency is equal to that of degenerate band edge. The initial rate of growth
of the frozen mode amplitude is independent of $N$ and described by
(\ref{FL DBE}). The limiting case (d) of the semi-infinite structure is also
shown in Fig. \ref{Amn6w}(c). In all cases, the incident wave has the same
polarization and unity amplitude.}
\label{SMNnd}
\end{figure}

\section{Conclusion}

In this paper we outlined several different manifestations of the frozen mode
regime in photonic crystals. Although all our numerical examples relate to
periodic layered structures, in fact, the frozen mode regime is more like a
universal wave phenomenon. Indeed, we can talk about different kinds of wave
excitations in low-loss periodic media. But as soon as the respective Bloch
dispersion relation displays a singularity like a stationary inflection point
(\ref{A SIP}) or a degenerate band edge (\ref{A DBE}), we have every reason to
expect a very similar behavior involving the frozen mode regime. In other
words, the possibility of the frozen mode regime is determined by some
fundamental spectral properties of the periodic structure. If a periodic array
is relatively simple, for instance, a stratified medium with one dimensional
periodicity, its frequency spectrum may prove to be too simple to support the
proper spectral singularity. The more complex the periodic structure is, the
more likely it will be capable of supporting such a phenomenon. For instance,
in the case of layered arrays we need birefringent layers and, sometimes, at
least three layers in a unit cell.

Another important question is how robust the frozen mode regime is. For
instance, what happens if we introduce a small absorption or structural
imperfections? Of course, these factors suppress the frozen mode amplitude.
But in this respect, the frozen mode regime is no different from any other
coherent or resonance effects it periodic structures. This problem can be
solved at any particular frequency range by appropriate choice of the
constitutive materials.

Another fundamental restriction relates to the size of the periodic structure.
In this paper we assumed that the periodic array occupies the entire
half-space $z\geq0$. A good insight on what happens to the frozen mode in a
finite periodic array is given by Fig. \ref{SMNnd}. These graphs prove that
the frozen mode regime in a finite periodic array can be as robust as that in
an imaginary semi-infinite structure. The optimal number of layers depends on
such factors as the absorption characteristics of the constitutive materials,
the geometrical imperfections of the periodic array, the desired degree of
field enhancement in the frozen mode, etc. On the other hand, in finite
(bounded) photonic crystals, some new resonance phenomena can arise, such as
transmission band edge resonances \cite{Strat1,Strat3,PRE 05}. These effects,
though, occur at slightly different frequencies and are qualitatively
different from the frozen mode regime.

\textbf{Acknowledgment and Disclaimer:} Effort of A. Figotin and I. Vitebskiy
is sponsored by the Air Force Office of Scientific Research, Air Force
Materials Command, USAF, under grant number FA9550-04-1-0359.

\end{document}